\documentclass{ws-mpla}

\usepackage[super]{cite}
\usepackage{xcolor}
\usepackage[verbose,hypertexnames=false]{hyperref}
\hypersetup{colorlinks=false,allbordercolors=blue,pdfborderstyle={/S/U/W 1}}

\def\be{\begin{equation}}
\def\ee{\end{equation}}
\def\bea{\begin{eqnarray}}
\def\eea{\end{eqnarray}}

\begin{document}

\markboth{Francisco S.N. Lobo and Tiberiu Harko}{Gravitational properties of Bose-Einstein condensate dark matter halos}

\catchline{}{}{}{}{}

\title{Gravitational properties of Bose-Einstein condensate dark matter halos}

\author{Francisco S.N. Lobo}

\address{Instituto de Astrof\'{i}sica e Ci\^{e}ncias do Espa\c{c}o, Faculdade de Ci\^encias da Universidade de Lisboa, Edif\'{i}cio C8, Campo Grande, P-1749-016, Lisbon, Portugal,\\
	Departamento de F\'{i}sica, Faculdade de Ci\^{e}ncias, Universidade de Lisboa, Edifício C8, Campo Grande, PT1749-016 Lisbon, Portugal\\
fslobo@fc.ul.pt}

\author{Tiberiu Harko}

\address{Faculty of Physics, Babe\c s-Bolyai University, 1 Kog\u alniceanu Street,\\ 400084 Cluj-Napoca, Romania,\\
	Astronomical Observatory, 19 Cire\c silor Street, 400487, Cluj-Napoca, Romania
}

\maketitle


\begin{abstract}
	Recent studies suggest that dark matter could take the form of a Bose--Einstein condensate (BEC), a possibility motivated by anomalies in galactic rotation curves and the missing mass problem in galaxy clusters. We investigate the astrophysical properties of BEC dark matter halos and their potential observational signatures distinguishing them from alternative models. In this framework, dark matter behaves as a self-gravitating Newtonian fluid with a polytropic equation of state of index $n=1$. We derive analytic expressions for the mass distribution, gravitational potential, and dynamical profiles such as the density slope and tangential velocity. The lensing behavior of BEC halos is analyzed, yielding a general series representation of the projected surface density that enables precise predictions for deflection angles, lensing potentials, and magnifications. Finally, halo equilibrium and stability are examined via the scalar and tensor virial theorems, leading to perturbation equations that describe their response to small disturbances. Together, these results provide a unified framework linking the microscopic physics of condensate dark matter to macroscopic halo observables.
\keywords{Bose--Einstein condensate; dark matter halos; gravitational properties.}
\end{abstract}


\section{Introduction}

A key result of quantum statistical mechanics is that, at sufficiently low temperatures, a dilute gas of identical bosons undergoes a collective phase transition in which all particles condense into the ground state, forming a \textit{Bose--Einstein condensate} (BEC) \cite{rev3,rev7}. First predicted by Bose and Einstein, and realized experimentally many decades later \cite{exp1,rev2}, the BEC represents a macroscopic manifestation of quantum mechanics on mesoscopic and even macroscopic scales.  
The transition occurs once the thermal de Broglie wavelength, $\lambda_{T} = \sqrt{2\pi \hbar^{2}/m_\chi k_{B} T}$,
becomes comparable to the mean interparticle spacing $l \sim n^{-1/3}$. Here $m_\chi$ is the particle mass, $T$ is the temperature, $n$ is the number density, $\hbar$ is the reduced Planck constant, and $k_{B}$ is Boltzmann’s constant. In this regime, particle wave functions strongly overlap, and the system behaves as a coherent quantum fluid.  
The critical temperature for condensation follows from the condition $\lambda_{T} > l$, yielding the scaling relation: $T < \left(\frac{2\pi \hbar^{2}}{m_\chi k_{B}}\right) n^{2/3}$,
which explicitly connects the onset of condensation to the microscopic parameters of the system \cite{rev3,rev7}. This result underpins the study of ultra-cold atomic gases and has broad implications in condensed matter, cosmology, and astrophysics, where BEC-like states are invoked in models of dark matter and compact objects.

A coherent quantum condensate forms under two conditions: either at high particle densities, where wave functions overlap even at moderate temperatures, or at ultra-low temperatures, where the thermal de Broglie wavelength spans many particles. In both scenarios, this overlap generates long-range quantum coherence, the defining feature of a Bose--Einstein condensate (BEC). The phenomenon was spectacularly confirmed in the 1990s, when experiments achieved quantum degeneracy in dilute atomic gases \cite{exp1,rev2}. By using a combination of laser cooling, magnetic trapping, and evaporative cooling, researchers reached nanokelvin temperatures, allowing a macroscopic fraction of atoms to occupy the ground state, and realizing the first controlled BECs.

These experimental advances confirmed fundamental principles of quantum statistical mechanics and opened new avenues across physics. In atomic physics, BECs allow studies of coherence in many-body systems; in statistical mechanics, they exemplify quantum phase transitions; and in condensed matter, they model superfluidity and superconductivity. Beyond the laboratory, BECs have inspired astrophysical applications, including the proposal that dark matter may exist as an ultra-light bosonic condensate \cite{BoHa07}, with implications for galactic halos, structure formation, and cosmic evolution. This possibility continues to drive research at the intersection of particle physics, cosmology, and condensed matter theory.

The polytropic gas model is a fundamental framework for describing self-gravitating systems, widely applied to compact astrophysical objects such as white dwarfs, neutron stars, and stellar cores. It assumes a simple power-law relation between pressure $P$ and energy density $\rho$,  $P = K \rho^{1+1/n}$,
where $K$ is a constant related to the entropy, and $n$ is the \textit{polytropic index}, which controls the stiffness of the equation of state, and thus the structural properties of the configuration: smaller $n$ corresponds to stiffer matter, larger $n$ to softer matter \cite{Hor}.  

The equilibrium structure of a spherically symmetric polytropic fluid follows from hydrostatic balance and the Poisson equation for gravity, which together reduce to the dimensionless \textit{Lane--Emden equation},
\begin{equation}
	\frac{1}{\zeta^{2}} \frac{d}{d\zeta}\!\left(\zeta^{2}\frac{d\theta}{d\zeta}\right) + \theta^{n} = 0,
	\label{eq:LaneEmden}
\end{equation}
where $\theta(\zeta)$ denotes the normalized density profile. The density is given by $\rho = \rho_{c}\,\theta^{n}$ with $\rho_{c}$ the central density, while the dimensionless radius is defined as $\zeta = \sqrt{\frac{4\pi G \rho_{c}^{2}}{(n+1)P_{c}}}\, r$,
with $P_{c}$ the central pressure and $G$ Newton’s constant. The entire configuration is thus determined by the solution $\theta(\zeta)$ of Eq.~(\ref{eq:LaneEmden}), subject to the regular boundary conditions $\theta(0)=1$ and $\theta'(0)=0$.

The Lane--Emden equation plays a central role in astrophysics and mathematical physics, and its properties have been extensively investigated through analytical and numerical methods. Despite its simple form, its nonlinear nature generally prevents closed-form solutions, so numerical integration is required for most polytropic indices $n$. Exact solutions exist only for three special cases: $n=0$, $n=1$, and $n=5$. The $n=0$ case corresponds to a uniform-density configuration, while $n=5$ describes an extended structure of infinite radius but finite mass, mainly of mathematical interest.  
The intermediate case $n=1$ admits the analytic solution
\begin{equation}
	\theta(\zeta) = \frac{\sin \zeta}{\zeta},
\end{equation}
which exhibits oscillatory behavior distinct from the other exact solutions. Although this profile was initially regarded as physically unrealistic for compact stars, later developments revealed its fundamental significance in an unexpected domain: the physics of dilute Bose--Einstein condensates. In the mean-field description of weakly interacting bosons at ultra-low temperatures, the effective equation of state of a BEC is precisely polytropic with index $n=1$. This striking connection between stellar astrophysics and quantum fluids has revived interest in the $n=1$ Lane--Emden solution, and opened new avenues for both astrophysical modeling and condensed matter physics.

These aspects will be reviewed in this article, which is based on Ref. \cite{Harko:2015aya}.

\section{Bose-Einstein condensate dark matter halos}

\subsection{General considerations}

We adopt the working hypothesis that galactic dark matter halos consist of a strongly coupled, cold, and dilute Bose--Einstein condensate (BEC) at zero temperature. This framework provides a physically motivated description of large-scale structure and halo dynamics, while bridging insights from condensed matter physics and astrophysics.  
In this regime, interactions between dark matter particles are dominated by low-energy binary collisions. The detailed form of the two-body potential is irrelevant; instead, the physics is captured by a single parameter, the \textit{$s$-wave scattering length} $l_{a}$, which characterizes the strength of interactions and determines the macroscopic properties of the condensate halo \cite{rev2}.

Assuming that DM consists of a BEC, rotating with an angular velocity $\omega$, it must then be described by the coupled Gross-Pitaevskii-Poisson system. The nonlinear Gross-Pitaevskii equation in the presence of an external potential $V$ has the following form \cite{rev3,rev7}
\begin{equation}
	i \hbar \frac{\partial \psi}{\partial t}=-\frac{\hbar^2}{2m_\chi}\nabla^2\psi +\psi \int{|\psi \left(r',t \right)|^2 U_I \left(|-r'| \right)}dr'+\left(V+V_{rot}\right)\psi,
\end{equation}
  where $V_{rot}=-\omega ^2r^2$ is the rotational potential,  $U_I \left(|r-r'| \right)$ is the microscopic interaction potential between bosons, and  $m_{\chi}$ is the dark matter particle mass. 
A standard approximation in the theory of dilute Bose gases is to replace $U_I \left(|r-r'| \right)$ with an effective contact interaction
$	U_{I}(\vec{r}' - \vec{r}) = U_{0}\, \delta(\vec{r}' - \vec{r})$,
where $U_{0}$ is an effective coupling constant and $\delta$ denotes the Dirac delta distribution. $U_{0}$ is related to the $s$-wave scattering length $l_{a}$ by
\begin{equation}
	U_{0} = \frac{4\pi \hbar^{2} l_{a}}{m_{\chi}},
\end{equation}

The scattering length $l_a$ is related to the scattering cross section $\sigma$ by the relation $\sigma =\left(4\pi/{\tilde k}^2\right)\delta _0^2=4\pi l_a^2$, with ${\tilde k}$ the wave vector of the scattered wave, while $\delta _0=-{\tilde k}l_a$ \cite{rev3}. This simplification renders the many-body Hamiltonian tractable, while retaining the essential physics of short-range interactions. Moreover, it leads to the Gross-Pitaevskii equation \cite{BoHa07}
\begin{equation}\label{GP}
	i \hbar \frac{\partial \psi}{\partial t}=-\frac{\hbar^2}{2m}\nabla^2\psi +V_{rot}\psi+ U_0 |\psi|^2 \psi+V\psi.
\end{equation}

Thus, the condensate dynamics is fully characterized by the pair $(m_{\chi}, l_{a})$, which determine the macroscopic behavior of BEC dark matter halos. This effective field-theoretic description, rooted in ultra-cold atomic physics, provides a natural starting point for studying their structure and stability.

For the potential $V$ we assume it is the gravitational potential $V_{grav}(\vec{r})$, $V(\vec{r})=V_{grav}(\vec{r})$, and it is determined self-consistently via the Poisson equation,
\begin{equation} \label{poi}
	\nabla^{2} V_{grav}(\vec{r}) = 4 \pi G \rho(\vec{r}),
\end{equation}
with mass density 
\begin{equation}
	\rho(\vec{r}) = m_{\chi} |\psi(\vec{r})|^{2}.
\end{equation}
The total particle number is fixed by the normalization condition
$	N = \int |\psi(\vec{r})|^{2} \, d^{3}\vec{r}$.
ensuring particle number conservation throughout the halo.

Eqs.~(\ref{GP})--(\ref{poi}) together define the Gross--Pitaevskii--Poisson (GPP) system, a set of coupled nonlinear equations that form the foundation for modeling Bose--Einstein condensed dark matter halos. The GPP framework enables the study of equilibrium structure, stability, and macroscopic dynamics, providing a natural link between the microscopic physics of quantum condensates and the astrophysical scales relevant for dark matter.

The study of the GPP system is significantly simplified through the use of the hydrodynamic representation,  which is obtained by representing $\psi$ in the form
$
\psi\left(\vec{r},t\right)=\sqrt{n\left(\vec{r},t\right)}e^{iS\left(\vec{r},t\right)/\hbar}
$, where $S\left(\vec{r},t\right)/\hbar$ is the phase of the wave function. Then the Gross-Pitaevskii equation (\ref{GP}) can be reformulated as a continuity and a hydrodynamic-type Euler equation, which are given by
\be\label{cont1}
\frac{\partial n}{\partial t}+\nabla \cdot \left(n\vec{v}\right)=0,
\ee
and
\bea\label{eu1}
m\frac{d\vec{v}}{dt}&=&m\left[\frac{\partial \vec{v}}{\partial t}+\left(\vec{v}\cdot \nabla\right)\vec{v}\right]=-\nabla \left[V\left(\vec{r}\right)+V_{rot}\left(\vec{r}\right)+
U_0n-\frac{\hbar ^2}{2m\sqrt{n}}\Delta \sqrt{n}\,\right],
\eea
respectively. Here $\vec{v}$ is the velocity of the dark matter condensate, defined according to $\vec{v}=\nabla S\left(\vec{r},t\right)/m$. The last term in Eq.~(\ref{eu1}) is the quantum potential. 

The condensate pressure is determined by the short-range interactions and is given by
\begin{equation} \label{pressn}
	p = \frac{U_{0}}{2 m_{\chi}} \rho^{2} = \frac{2 \pi \hbar^{2} l_{a}}{m_{\chi}^{2}} \rho^{2}.
\end{equation}

Together, Eqs.~(\ref{cont1})--(\ref{pressn}) show that the condensate behaves as a quantum fluid obeying hydrodynamic-like conservation laws, with an effective equation of state $p \propto \rho^{2}$. This quadratic dependence corresponds to a polytropic equation of state with index $n=1$, directly linking the microscopic properties of the Bose--Einstein condensate to the macroscopic structure of self-gravitating dark matter halos.

For a static dark matter halo  all terms containing time derivatives cancel, and $\vec{v}\equiv 0$. In the limit of a large number of particles, the quantum pressure contribution to the total energy density is negligible throughout most of the condensate, becoming relevant only near its boundary. This observation motivates the \textit{Thomas--Fermi approximation}, in which the quantum pressure term in the Gross--Pitaevskii equation is neglected. The approximation becomes increasingly accurate as $N$ grows, and is formally exact in the limit $N \to \infty$.   Hence  in the Thomas-Fermi approximation for a static dark matter halo Eq.~(\ref{eu1}) takes the form
\be\label{13}
 \nabla \left[U\left(\vec{r}\right)+V_{rot}\left(\vec{r}\right)+U_0n\right]=0.
\ee
From Eq.~(\ref{13}), after applying again the $\nabla$ operator, it follows that the basic equation describing the rotating BEC dark matter halos is given by
\be\label{dens}
\Delta \rho \left(\vec{r}\right)+k^2\left[\rho \left(\vec{r}\right)-\frac{\omega ^2}{2\pi G}\right]=0,
\ee
where we have denoted
\be
k^2=\frac{4\pi Gm_\chi^2}{U_0}=\frac{Gm_\chi^3}{l_a\hbar ^2}.
\ee
The last term in Eq.~(\ref{dens}) accounts for the centrifugal effect in the $x$--$y$ plane. For static condensates, $\omega = 0$, and we obtain
\begin{equation} \label{eqpoi}
	\nabla^{2} \rho + k^{2} \rho = 0
\end{equation}
where $k$ encodes the balance between gravitational attraction and repulsive two-body interactions. Equation~(\ref{eqpoi}) is a Helmholtz type equation, providing analytic solutions for static BEC dark matter halo density profiles and illustrating the connection between microscopic interaction parameters and the large-scale structure of self-gravitating condensates.

Thus, Eqs.~(\ref{cont1}) and (\ref{eu1}) illustrate how the Gross--Pitaevskii formalism, under the Thomas--Fermi approximation, reduces to a quantum hydrodynamic description. This framework closely parallels classical fluid dynamics, while retaining its quantum origin, providing a powerful tool for analyzing the equilibrium, rotation, and stability of Bose--Einstein condensed dark matter halos.

\subsection{Mass and gravitational properties of static condensed dark matter halos}

The general solution of Eq.~(\ref{eqpoi}) for a static, spherically symmetric Bose--Einstein condensate dark matter halo is \cite{BoHa07}
\begin{equation} \label{neq1}
	\rho(r) = \rho_{c} \, \frac{\sin (k r)}{k r},
\end{equation}
where $\rho_{c}$ is the central density. This profile corresponds to a polytrope with index $n=1$, directly linking the microscopic condensate physics to the macroscopic halo structure.  

The density vanishes at the boundary $r = R$, yielding $k R = \pi$. Using $k = \sqrt{4 \pi G m_{\chi}^{2}/U_{0}}$, the halo radius can be expressed as
\begin{equation} \label{rad}
	R = \frac{\pi}{k} = \pi \sqrt{\frac{\hbar^{2} l_{a}}{G m_{\chi}^{3}}}=\pi\sqrt{\frac{\hbar ^2}{Gm_\chi^3}}\left(\frac{\sigma}{4\pi}\right)^{1/4},
\end{equation}
showing that the macroscopic size of the condensate halo is entirely determined by the particle mass $m_{\chi}$ and $s$-wave scattering length $l_{a}$. This establishes the $n=1$ polytropic model as a natural framework for describing equilibrium Bose--Einstein condensed dark matter halos.

The cumulative mass profile is
\begin{eqnarray}
	m(r) &=& 4 \pi \int_{0}^{r} \rho(r') \, r'^{2} \, dr' 
	= \frac{4 \pi \rho_{c}}{k^{2}} r \left[ \frac{\sin(k r)}{k r} - \cos(k r) \right],
\end{eqnarray}
where $R$ is the halo radius from Eq.~(\ref{rad}). This relation links the mass profile directly to the local density and slope.

The total mass of the halo at $r = R$ is
\begin{equation} \label{26}
	M(R) = \frac{4 \pi^{2} \rho_{c}}{k^{3}} = \frac{4}{\pi} \rho_{c} R^{3},
\end{equation}
showing a cubic scaling between mass and radius, determined by the central density. This simple relation connects the microscopic properties of the condensate to the halo’s global astrophysical characteristics.

The halo radius and total mass are determined by the particle mass $m_{\chi}$ and $s$-wave scattering length $l_a$. From Eq.~(\ref{rad}) we obtain the mass of the dark matter particle as
\begin{equation}\label{mdm}
	m_{\chi} = \left( \frac{\pi^{2} \hbar^{2} l_a}{G R^{2}} \right)^{1/3} 
	\approx 6.73 \times 10^{-2} \, [l_a (\mathrm{fm})]^{1/3} [R (\mathrm{kpc})]^{-2/3} \ \mathrm{eV},
\end{equation}
so that typical values ($l_a \approx 1$ fm, $R \approx 10$ kpc) yield $m_{\chi} \approx 14$ meV, while larger scattering lengths ($l_a \approx 10^6$ fm) correspond to $m_{\chi} \approx 1.44$ eV, linking laboratory and astrophysical scales. 

The dark matter-dark matter self-interaction cross section can be obtained from the study of the collisions between clusters of galaxies, like the Bullet Cluster (1E 0657-56) and the baby Bullet Cluster (MACSJ0025-12) \cite{Bul, Bul1}. If the ratio $\sigma _m=\sigma /m_{\chi}$  is known from observations, with the use of Eq.~(\ref{rad}) we obtain some limits on the mass of the dark matter particle.  

From  weak lensing, strong lensing, $X$-ray and optical observations of the merging of galaxy cluster 1E 0657-56, an upper limit (68 \% confidence) for $\sigma _m$ of the order of $\sigma _m<1.25\;{\rm cm^2/g}$ can be obtained \cite{Bul}. Hence we obtain for the mass of the dark matter particle the upper limit
\begin{eqnarray}
m_{\chi }<0.1791\times\left(\frac{R}{10\;{\rm kpc}}\right)^{-4/5} \times
\left(\frac{\sigma _m}{1.25\;{\rm cm^2/g}}\right)^{1/5}\;{\rm meV}.
\end{eqnarray}
 For the scattering length $l_a$, we obtain the constraint
\bea
l_a<\sqrt{\frac{\sigma _m\times m_{\chi }}{4\pi }}=1.7827\times 10^{-19}\;{\rm cm}=
1.7827\times 10^{-6}\;{\rm fm}.
\eea

A stronger constraint, $\sigma _m\in(0.00335\;{\rm cm^2/g},0.0559\;{\rm cm^2/g})$ was proposed in \cite{Bul1}, leading to 
\begin{equation*}
m_{\chi }\approx 
\left(0.053-0.093\right)\times \left(\frac{R}{10\;{\rm kpc}}\right)^{-4/5}\;{\rm meV}, \quad l_a\approx \left(5.038-27.255\right)\times 10^{-8}\;{\rm fm}.
\end{equation*}
Thus the combination of the galactic radii data and the Bullet Cluster constraints indicate a dark matter particle mass of the order of $m_{\chi }\approx 0.1$ meV.

\section{Lensing properties of condensed dark matter halos}

A key observational test of the Bose--Einstein condensate (BEC) dark matter model is provided by gravitational lensing \cite{Mielke1}, particularly the deflection of light by galactic halos. Photons traversing regions with approximately flat rotation curves probe the halo mass distribution and its underlying dynamics \cite{Sch}.  

The projected surface mass density, obtained by integrating the three-dimensional BEC density profile along the line of sight, plays a central role in determining the lensing properties. 

The surface mass density of a lens can be obtained as  $\Sigma \left( \xi \right) =\int_{-\infty }^{+\infty }\rho \left(
\xi ,r\right) dz$, where $\xi $ is the radius measured from the center of
the lens and $r=\sqrt{\xi ^{2}+z^{2}}$. It can be also written as an  Abel transform \cite{Sch}
\begin{equation}
\Sigma \left( \xi \right) =2\int_{\xi }^{+\infty }\frac{\rho \left( r\right)
rdr}{\sqrt{r^{2}-\xi ^{2}}}=2\rho _{c}\frac{R}{\pi} \int_{\xi }^{R}\frac{\sin
\left( kr\right) dr}{\sqrt{r^{2}-\xi ^{2}}}.
\end{equation}

The central surface mass density can be approximated as
\begin{equation}
\frac{\Sigma (0)}{2\rho _c}\approx R\left(1 -\frac{\pi ^2}{18}+\frac{\pi ^4}{%
600}-\frac{\pi ^6}{35280}\right)=0.5867R.
\end{equation}

Using the standard lens equation, the deflection angle of photons depends directly on the BEC density profile, allowing the model to be tested against observations \cite{Harko:2015aya}. Additionally, the magnification of background sources is determined by the second derivatives of the lensing potential, which in turn depend on the surface mass density. Comparing measured magnifications with predictions from the BEC profile provides an independent observational test of the model. 

\section{The virial theorem}

The virial theorem is a fundamental relation in theoretical physics that links the average kinetic energy of a system to its potential energy, providing a powerful method to study equilibrium and dynamical properties without solving the full microscopic equations \cite{Chand1}. Originally formulated for self-gravitating, rotating fluids, it allows investigation of equilibrium configurations and stability under small perturbations across a wide range of astrophysical and condensed matter systems.

A key generalization is the tensor virial theorem, which transforms the local Euler equations into global tensorial relations. These encode information about the overall structure, shape, and stability of a gravitating system, including rotational support, deformation, and energy balance. A classic application is the study of small perturbations of incompressible, uniform-density ellipsoids, where the virial equations describe normal modes of oscillation, providing insight into the system's dynamical response in the absence of viscous dissipation.

The scalar virial theorem can be obtained by studying the behavior of the physical parameters under the scaling transformation $\vec{r}\rightarrow \alpha \vec{r}$, where $\alpha $ is a
constant. From the normalization condition $N=\int \left| \psi \left( \vec{r}\right) \right| ^{2}d^3\vec{r}$ one obtains  $\psi \left(\vec{r}\right)\rightarrow \alpha ^{-3/2} \psi \left(\vec{r}\right)$. Hence the total energy behaves as
\begin{equation}
E\left[\alpha \right]=\alpha ^{-2}E_K+\alpha ^2E_{rot}+\alpha ^{-3}E_{int}+\alpha ^{-1}E_{grav}.
\end{equation}
Imposing the condition $\left.\left( \delta E\left[ \alpha \right] /\delta \alpha \right) \right|
_{\alpha =1}=0$, one obtains the virial theorem as \cite{Harko:2015aya}
\begin{equation}
2E_K-2E_{rot}+3E_{int}+E_{grav}=0,
\end{equation}
leading to
\begin{equation}  \label{75}
\mu N=E_{rot}+2E_{int}+2E_{grav},
\end{equation}
where $\mu$ is the chemical potential, and $E_{int}=\left(U_0/2\right)\int_V{\rho ^2(\vec{r})d^3\vec{r}}$, and $E_{grav}=\left(m_{\chi}^2/2\right)\int_V{V_{grav}\left(\vec{r}\right)\rho \left(\vec{r}\right)d^3\vec{r}}$, respectively.
The total energy of the dark matter halo can be obtained as
\begin{equation}
E=\frac{1}{2}\left[\frac{U_0}{m_{\chi }}\rho _c+m_{\chi }V_{grav}(0)\right]+\frac{1}{2}E_{rot}.
\end{equation}

By using the density for the static condensate given by Eq.~(\ref{neq1}), it follows that $E_K\approx \bar 2M(R)k^2/m_{\chi}^2$ and $E_{int}\approx
U_0M(R)^2k^3/m_{\chi }^2$, respectively. The Thomas-Fermi approximation requires $E_K \ll E_{int}$, giving
\begin{equation}
N=\frac{M}{m_{\chi }} \gg \frac{1}{kl_a}=\frac{R}{\pi l_a}, \qquad R \gg \sqrt{\frac{%
m_{\chi }}{4l_{a}\rho _{c}}}=\sqrt{\frac{1}{4l_an_c}}.
\end{equation}
Hence for systems with enough high particle numbers $n_c$, the Thomas-Fermi
approximation is always valid.

Alternatively, the scalar virial theorem for the Bose--Einstein condensate dark matter halos allows one to assess the validity of the Thomas--Fermi approximation by comparing kinetic and interaction energies, which yields a lower bound on the halo radius
\begin{equation}
	R \gg 1.581 \times 10^{3} \, 
	\left( \frac{m_{\chi}}{10^{-37}\ \mathrm{g}} \right)^{1/2} 
	\left( \frac{l_{a}}{10^{-20}\ \mathrm{cm}} \right)^{-1/2} 
	\left( \frac{\rho_{c}}{10^{-24}\ \mathrm{g/cm^{3}}} \right)^{-1/2} \ \mathrm{cm},
\end{equation}
ensuring that the quantum pressure is negligible compared to interaction energy.

The stability of the BEC dark matter halos can be studied by using the virial theorem. Assuming that the evolution of the initial coordinates $a_i$ is a linear expression,
$x_i=a_i\zeta (t)$ with $i=1,2,3$, the equation governing the time evolution of $\zeta (t)$ is given by \cite{Harko:2015aya}
\begin{equation}  \label{117}
\frac{d^2\zeta }{dt^2}=\frac{1}{\zeta ^2}\left(\frac{1}{\zeta ^2}-1\right)%
\frac{\left |\Phi _0\right|}{I_0}+\frac{1}{\zeta ^3}\left(1-\frac{1}{\zeta }%
\right)\frac{2K_0}{I_0},
\end{equation}
where $\Phi _0$, $K_0$ and $I_0$ correspond to the rotating dark matter halo.  In the first
approximation, $\zeta (t)$ can be represented as
$
\zeta (t)=1+\epsilon (t)
$,
where $\epsilon (t)\ll 1$. Hence Eq.~(\ref{117}) takes the form 
\begin{equation}  \label{119}
\frac{d^2\epsilon}{dt^2}+\sigma ^2\epsilon =\lambda \epsilon ^2+O(\epsilon
^3),
\end{equation}
where 
\begin{equation}  \label{122}
\sigma ^2=2\frac{\left |\Phi _0\right |}{I_0}-\frac{2K_0}{I_0}, \qquad \lambda =\frac{1}{2}\left(7\sigma ^2-\frac{2K_0}{I_0}\right).
\end{equation}

Equations~(\ref{122}) provide the stability conditions for BEC dark matter halos in the linear approximation. If the halo is initially non-rotating, we have $%
K_0=0$, and the stability condition reduces to
$
\sigma ^2=2\frac{\left |\Phi _0\right |}{I_0}>0
$,
a condition which is always satisfied by BEC dark matter halos.
The initial gravitational energy of the dark matter halo is $\left |\Phi _0\right |=(3/4)GM^2/R$,
and the moment of inertia is $I_0=2MR^2/5$. The oscillation frequency of the halo is given by
$\sigma ^2=\frac{15GM}{8R^3}=\frac{15G}{2\pi}\rho _c$, and the period of the oscillations is
\begin{eqnarray}
T=\frac{2\pi}{\sigma }&=&\sqrt{\frac{8}{15}}\pi ^{3/2}\frac{1}{\sqrt{G\rho _c%
}}=1.5745\times 10^{16}\times  \left(\frac{\rho _c}{10^{-24}\;\mathrm{g/cm^3}}\right)\;\mathrm{s}.
\end{eqnarray}

\subsection{Summary and Conclusion}

In the Bose--Einstein condensation model, dark matter is described as a non-relativistic, gravitationally confined Newtonian gas with a polytropic equation of state of index $n=1$. This framework connects the microscopic properties of dark matter particles to the macroscopic structure of galactic halos.

The mass distribution and gravitational properties of the condensate halos have been derived explicitly, yielding analytic expressions for the density profile, cumulative mass, and gravitational potential, allowing detailed modeling of halo structure within the BEC paradigm. Gravitational lensing properties, including deflection angles, magnification factors, and surface mass densities, can be computed precisely, enabling direct comparison with observational data.

The stability and dynamical behavior of BEC halos can be analyzed rigorously via the scalar and tensor virial theorems, providing insight into equilibrium configurations, oscillation modes, and the response to small perturbations. Overall, Bose--Einstein condensates offer a compelling dark matter candidate, combining a well-defined microscopic framework with predictive macroscopic properties and testable observational consequences, motivating further theoretical and observational study.

\section*{Acknowledgments}
FSNL acknowledges support from the Funda\c{c}\~{a}o para a Ci\^{e}ncia e a Tecnologia (FCT) research grants UIDB/04434/2020, UIDP/04434/2020 and PTDC/FIS-AST/0054/2021, and from the FCT Scientific Employment Stimulus contract with reference CEECINST/00032/2018.

\section*{ORCID}

\noindent Francisco S.N. Lobo - \url{https://orcid.org/0000-0002-9388-8373}

\noindent Tiberiu Harko - \url{https://orcid.org/0000-0002-1990-9172}


\end{document}